\documentstyle[prb,aps]{revtex}
\begin{document}
\author{V.N. Soshnikov}
\address{Krasnodarskaya St. 51-2-168, 109559 Moscow, Russia,\\
Department of Physics, Institute of Scientific and Technical Information of\\
the Russian Academy of Sciences (VINITI, 125219 Moscow Usievitch St. 20 A)%
\thanks{%
Now retired.}.}
\title{A NEW LOOK AT THE LANDAU'S THEORY OF SPREADING AND DAMPING OF WAVES IN
COLLISIONLESS PLASMAS\thanks{%
This paper is a survey of development of results, published in difficult to
access for a foreign reader issues of the Institute of Scientific and
Technical Information of the Russian Academy of Sciences (VINITI): 125219
Moscow Usievitch St. 20A, Department of Physics (V.N. Soshnikov, Deponents
VINITI 6480-B 88 on July 22, 1988: n 3361-B 91 on July 23, 1991, and n
2982-B 90 on May 16, 1990. There are also references to related foregoing
author's publications VINITI, including investigations on the low collision
approach).}}
\maketitle

\begin{abstract}
The theory of plasma waves and Landau damping in Maxwellian plasmas,
Landau's ``rule of pass around poles'' include doubtful statements,
particularly related to an artificial ``constructing'' of the dispersion
equation, what should allow the possibility of its solution otherwise not
existing at all, and the possibility of analytical continuations of
corresponding very specific ruptured functions in the one-dimensional
Laplace transformation, used by Landau, what is the base of his theory.

We represent, as an accessible variant, a more general alternative theory
based on a two-dimensional Laplace transformation, leading to an
asymptotical in time and space solution as a complicated superposition of
coupled damping and {\em non-damping \/} plane waves and oscillations with
different dispersion laws for every constituent mode. This theory naturally
and very simply explains paradoxes of the phenomenon of plasma echo. We
propose for discussion a new ideology of plasma waves (both electron and
ion-acoustic waves) qualitatively different from the traditional theory of
Landau damping for non-collisional as well as for low-collisional plasmas.
\end{abstract}


\tighten


The sophisticated theory of Landau damping of longitudinal plasma waves,
which was published by Landau in 1945 [1], is now considered as the greatest
success of the physics of plasma, the heritage verified experimentally and
theoretically, which has not to be doubted. But this halo is to a
considerable extent dispersed after some critical analysis of different
sides of the problem.

1. According to the classical linear theory of longitudinal plasma waves, in
order to derive the dispersion equation one has to substitute expressions of
the type of $\exp (-i\omega t+ikx)$ for electrical field $E$ and
perturbation $f_1(\vec v,x,t)$ of the isotropic electron distribution
function $f_0(v)$ into the linearized non-collisional kinetic equation

\begin{equation}
\frac{\partial f_1}{\partial t}+v_x\frac{\partial f_1}{\partial x}+\frac{eE_x%
}{m_e}\frac{\partial f_1}{\partial v_x}=0,
\end{equation}

\noindent and Poisson's equation

\begin{equation}
\frac{\partial E_x}{\partial x}=4\pi e\int_{-\infty }^\infty f_1d\vec{v},
\end{equation}
where $E_x$ is the selfconsistent electrical field in $x$ direction, $v_x$
is the $x$-component of the electron velocity, $e,m_e$ are correspondingly
the electron charge and mass, $\omega $ and $k$ are generally complex; and
then to solve the resulting dispersion equation $D(\omega ,k)=0$ defining a
dependence $k=f(\omega )$ for electron plasma waves, one or more, according
to the number of its roots.

For plasma waves, as it can be immediately shown assuming $\omega =\omega
_0-i\delta $ at real $\omega _0,\delta $ and $k$, breaking up integrals
according to velocity components in the range $-\infty <\vec{v}_i<\infty $
to constituents $-\infty <\vec{v}_x<0$ and $0<\vec{v}_x<\infty $ and picking
out the real and imaginary parts, after some very simple procedures: the
dispersion equation has no solutions for $\delta $. So, one obtains easily
for $D(\omega ,k)$ and imaginary part $%
\mathop{\rm Im}
D(\omega ,k)=0$, determining preferably $\delta $, equations
\[
D(\omega ,k)=1-\frac{4\pi ie^2}{km}\int_{-\infty }^\infty \frac{\partial
f_0(v)}{\partial v_x}\frac{d\vec{v}}{p+ikv_x}\equiv 1-iF(p),
\]

\begin{equation}
\mathop{\rm Im}
D(\omega ,k)=\frac{16\pi e^2\omega _0\delta }{m_e}\int_{-\infty }^\infty
dv_y\int_{-\infty }^\infty dv_z\int_{-\infty }^\infty \frac{\partial f_0}{%
\partial v_x}\frac{v_xdv_x}{(\omega _0^2-k^2v_x^2-\delta ^2)^2+4\delta
^2\omega _0^2},
\end{equation}

\noindent where $D(\omega ,k)$ is susceptibility, $F(p)$ is a particular
case of ``Landau function'', which will be discussed later, $p=-i\omega $,
and  $\partial f_0/\partial v_x<0$. Since near the point $v_x\simeq \omega
_0/k$ the integral appears to diverge as $1/\delta $, the whole expression
for $%
\mathop{\rm Im}
D(\omega ,k)$ seems to be finite, though ruptured at $\delta \rightarrow \pm
0$. But here the integrand in the left-hand side is sign-invariant for both
Maxwellian $f_0$ and any other $f_0$ with sign-invariatn $\partial
f_0/\partial v_x$, so this equation is evidently impossible to satisfy at
whatever $\omega _0\neq 0$ and any $\delta $, including $\delta \rightarrow
\pm 0$ (in contrary to $%
\mathop{\rm Re}
D(\omega ,k)=0$). Strongly in $\delta =0$ this expression makes no sense.
This equation has no solutions, too, independently on Landau theory and his
analytical continuations of such type of functions, on sophisticated Van
Kampen modes and whatever theories else.

The case of a low-collisonal plasma is analogous, but with replacing $\delta
$, in the simplest approximation, by $(\delta -\nu )$, where $\nu $ is
collision frequency, with the same result of non-existing solutions.

This statement is also equivalent to the note in [1] about non-existence of
complex poles in the right halfplane of the Laplace transformation used in
[1]. In the very strong canonical sense it means indeed non-existence of
solutions of the type of a single (or of some number, according to possible
roots at fixed $k$, as in multimode Landau's asymptotical solution)
monochromatic plasma waves.

One may also expand the terms of the dispersion equation, after integration
over velocity spherical angles $d\varphi d\theta $, in the small parameter $%
\delta /\omega _0\ll 1$, but the resulting series is asymptotically
divergent (e.g., cf. [2]), and according to the theory of such expansions,
the error of a resulting (in this approximation) dispersion equation
solution for $\delta $ (which just coincides here with the traditional
Landau's expression for $\delta $) turns out to be comparable in value with
the error, defined by a linear in $\delta $ term of the expansion in $\delta
$, i.e. the solution error is comparable with this solution itself.

2. Landau's the most slowly damping asymptotical wave should indeed satisfy
the dispersion equation (3). But the existence of a solution of the
disperation equation is attained by including in it some additional terms
defined according to considerations not related to the dispersion equation
inself. It is found that such ``artificially constructed'' (i.e. to the
Vlasov integral in the arbitrary sense of principal value one arbitrarily
adds the half-residium with semi-circle contour in the complex $v_x$-plane
in the sense of Landau's rule) dispersion equation which just now has
solutions, corresponds at asymptotically large times to plane attenuating
waves. But it is trivially evident that the dispersion relation does not
contain variables $x$ and $t$, and at $t\rightarrow \infty $ the precise
dispersion equation does not turn into the modified dispersion equation of
Landau theory.

The terms added by Landau [1] are found by a calculation of some contour
integral along the real axis in the plane of the complex variable $v_x$ with
passing around the pole $-i\delta /k$ near the point $\omega _0/k$ of the
real axis at $\delta \rightarrow +0$ (Fig.1).

\unitlength .1mm

\begin{picture}(100,180)(0,-140)
\put(40,0){\vector(4,0){60}}
\put(100,0){\line(10,0){60}}
\put(220,0){\vector(4,0){60}}
\put(280,0){\line(10,0){60}}
\put(190,0){\oval(60,60)[b]}
\put(190,0){\circle*{4}}
\put(190,-190){a}
\put(500,0){\vector(4,0){60}}
\put(560,0){\line(10,0){60}}
\put(640,0){\vector(4,0){60}}
\put(700,0){\line(10,0){60}}
\put(620,0){\line(0,-1){105}}
\put(640,0){\line(0,-1){105}}
\put(620,-130){\oval(50,50)[l]}
\put(640,-130){\oval(50,50)[r]}
\put(620,-155){\line(1,0){20}}
\put(630,-130){\circle*{4}}
\put(660,-80){$-i\delta$}
\put(630,-190){b}
\end{picture}

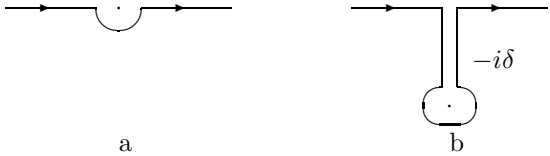
\vspace{0.25in} \indent {\bf Fig. 1 (a) and (b).} Some alternative variants
of calculation (with different results) of the Landau contour integrals $F(p)
$, $p=-i\omega $, in the plane of complex $v$ with $\omega \equiv \omega
_0-i\delta $ (real $\omega _0$, $\delta >0$) at $\delta \rightarrow +0$. %
\vspace{0.25in}

It is easy to see that in dependence on a way of tending $\delta $ to zero
there are equally accessible, as an example, both possibilities:  (a)
(half-residium in a pole near the real axis, what is a result of the Landau
rule of going around the pole), and (b) (total residium in the pole), as
well as other variants, including those leading to possible infinite
results. Any discussion about a preferable rightness of either the first or
the second result (see e.g. [3]) appears to be objectless because of
non-existing any definite limit of the contour integral at $\delta
\rightarrow 0$ at all. This means, that the result of this calculations
depends on a selected way of tending $\delta \rightarrow 0$ relative to the
contour integral calculation procedure in the complex plane $v_x$. Because
of it an analytical continuation of the regular in the upper halfplane of
complex $\omega \equiv ip$ function $F(p)$ into the lower halfplane $\omega $
appears to be non existing.

For calculation of expressions of the type (3) at $\delta \rightarrow \pm 0$
one might use the relation which is known in mathematics as Sokhotsky
formula
\begin{equation}
\lim_{\epsilon \rightarrow \pm 0}\int_{-\infty }^\infty \frac{f(x)dx}{%
x+i\epsilon }=\mp i\pi f(0)+{\rm v.p.}\int_{-\infty }^\infty \frac{f(x)dx}x
\end{equation}

\noindent which thus appears to be only a particular case because of the
limit existence being related to a selected definite procedure of
calculation of an integral (according to the definition) as a limit of the
corresponding sum
\begin{equation}
\int_{-\infty }^\infty \frac{f(x)dx}{x+i\epsilon }=\lim_{\Delta
x_k\rightarrow 0}\sum_k\frac{f(x_k)}{x_k+i\epsilon }.
\end{equation}

\noindent Substitution of this definition (5) into the left-hand side of eq.
(4) results in a double-limit sum, the value of which, as being a limit, is
evidently depending on the arbitrary choice of the relative cooperative
rates of tending to zero of the values of $\Delta x_k$ and $\epsilon $.

Strictly speaking, it takes plase also for the above considered expression
(3) $\lim [%
\mathop{\rm Im}
D(w,k)]$ (and $F(p)$) at $\delta \rightarrow \pm 0$. So, the above pointed
out result of some definite finite limits for that one is only some
particular way of calculation of the mentioned double-limit indefinite sum.
That is, the expression $\lim [%
\mathop{\rm Im}
D(w,k)]$ in (3) has really no definite sense at all.

3. In the well known paper of Landau [1] the field function $E(x,t)$ was
found by solving coupled equations (1) and (2) at an in advance arbitrarily
taken coordinate dependence $E(x,t)=\varphi (t)$.$exp(ikx)$ and $f_1\sim $ $%
exp(ikx)$ at all the times (cf. also [4],[5]) by means of the one-sided one
dimensional Laplace time transformation with the use of an analytical
continuation of a function of  the form
\begin{equation}
F(p)=\int_{-\infty }^\infty \frac{\psi (v)dv}{p+ikv}\equiv F(\omega ,k)
\end{equation}

\noindent from the right half-plane of complex $p$ to the left through the
peculiar line $%
\mathop{\rm Re}
p\equiv \epsilon =0$ and with the plane damping wave as an asymptotical
solution ($v$ is hereinafter $x$-component of velocity with omitted
subscript $x$; $\psi (v)$ is a real function of $v$).

Introducing complex $v$ with the counterclockwise rule of going around the
pole $v_p=ip/k$ at $%
\mathop{\rm Re}
p<0$ (Landau's rule) at integration by $v$, one removes the rupture at the
line $\epsilon \rightarrow \pm 0$ defined e.g., as some partial cases, by
the Sokhonsky formula (half-residium) or by the residue theorem, but does
not remove the rupture of derivatives $dF(p)/dp$ in the analyticity
Cauchy-Riemann conditions in the direction which is perpendicular to the
line $%
\mathop{\rm Re}
p=0$. So, at $%
\mathop{\rm Re}
p<0$ one finds, if, e.g. using the residue theorem in the calculation of the
contour integral (6), correspondingly at the left-hand side ($-$) and at the
right-hand side ($+$) from the line $%
\mathop{\rm Re}
p=0$:

\begin{equation}
F_{-}(p)={\rm v.p.}\int_{-\infty }^\infty \frac{\psi (v)dv}{p+ikv}+\frac{%
2\pi }k\psi (ip/k),\quad \qquad (\epsilon \leq 0)
\end{equation}

\noindent (that is the Landau's type analytical continuation to the left),
and

\begin{equation}
F_{+}(p)=\int_{-\infty }^\infty \frac{\psi (v)dv}{p+ikv},\qquad (\epsilon
>0).
\end{equation}

At the calculation of derivatives across (perpendicularly to) the line $%
\mathop{\rm Re}
p=0$ with a function $\psi $ having not peculiarities, one finds a finite
expression
\begin{equation}
\frac{dF_{-}(p)}{dp}=\frac{2\pi }k\frac d{dp}\psi (ip/k),\qquad (\epsilon
\rightarrow -0),
\end{equation}
$\qquad $

\noindent but contrary to it a divergent (infinite) at $\epsilon \rightarrow
+0$ integral expression (with infinite principal value at $\epsilon =0$) for
the derivative of the analytical at $\epsilon \neq 0$ function (8)
\begin{equation}
\frac{dF_{+}(p)}{dp}=-\int_{-\infty }^\infty \frac{\psi (v)dv}{(p+ikv)^2}%
,\qquad (\epsilon \rightarrow +0).
\end{equation}

4. To the very hard perceived consequences of the Landau theory one should
attribute also the existence of the pole $p=-ikv$, which appears at the
Landau procedure of analytical continuation of the Laplace transform of the
distribution function $f_1(v,x,t)$. This pole leads to an additive term for
the distribution function of the type $H(\vec{v})\exp (ikx-ikvt)$, which one
uses for an interpretation of the plasma echo [4]. Dramatic attempts to
comprehend and to conciliate this ``mysterious'' term with the common sense
are clearly demonstrated in the textbook [4] and are related to the
appearing paradoxicability of the existence of non-damping plasma
oscillations in the absence of whatever recovery force, whereas the electric
field had to be disappearing due to the Landau damping!

Such $f_1$, independent on $E(x,t)$, can not satisfy the Poisson equation
(2), since its substitution in the right hand side of (2) leads to quite
arbitrary functions of time and is not consistent with the field damping in
the left hand side. At the same time the paradox of plasma echo appears to
be naturally explained in principle in the further developed theory.

5. In analogy with [1] let us consider a semiinfinite (halfspace slab)
plasma with an {\sl initial} perturbation
\begin{equation}
f_1(v,x,t)=
{g(|\vec{v}|)\exp (ikx)
\text{ \qquad at }x\geq 0 \atopwithdelims\{.
0\qquad \qquad \qquad \qquad \text{at }x<0,}
\end{equation}

\noindent where $v$ is now the velocity component along $x$, and $g(|\bar{v}%
|)$ is some integrable function of $|\bar{v}|$.

Equation (1) is easily solved according to the known procedure with the
method of characteristics (by reduction to the equivalent system of simple
ordinary differential equations) with the result
\begin{equation}
f_1(v,x,t)=g(|v|)e^{ikx}+\int_{(x,x^{\prime }>0)}^tw[x-v(t-t^{\prime
}),t^{\prime }]dt^{\prime }
\end{equation}

\noindent where
\begin{eqnarray}
w(x,t) &\equiv &-\frac{eE(x,t)}{m_e}\frac{\partial f_0}{\partial v_x}; \\
x^{\prime } &\equiv &x-|v|(t-t^{\prime })
\end{eqnarray}

\noindent and an additional physical condition $x^{\prime }>0$ is related to
the finite speed of spreading of particles which are arriving to the point $%
x $.

Substituting expressions (12) - (14) into eq. (2) one obtains a linear
integro-differential equation for  $E(x,t)$:
\begin{equation}
\frac{\partial E(x,t)}{\partial x}=\frac{\sqrt{8\pi m_e}e^2n_e}{k_BT}%
\int_{-\infty }^\infty dvv\exp (-m_ev^2/2k_BT)\times
\int_{t_0(v)}^tdt^{\prime }E[x-v(t-t^{\prime }),t^{\prime }]-4\pi e\alpha
(x)\exp (ikx)
\end{equation}

\noindent where the dependence $\exp (ikx)$ is supposed only for the sake of
correspondence with the assumed by Landau coordinate dependence [1], and $%
\alpha (x)=\alpha _0=const$ at $x\geq 0$ ; $\alpha (x)=0$ at $x<0$ ; $%
t_0(v)=0$ at $t-x/|v|<0$ ; $t_0(v)=t-x/|v|$ at $t-x/|v|\geq 0$.

The one-value solution of eq. (15) must be defined in the same extent by
both initial and boundary conditions. The dependence $E(x,t)$ on $t$, as
well as on $x$ is defined by solving equation (15), and some analogy with
the Landau problem [1] for the infinite plasma may be realized by taking the
limit $x\rightarrow \infty $.

The general solution of equation (15) can be found with onesided
two-dimensional Laplace transformation:
\begin{eqnarray}
E(\zeta ,\tau ) &=&\frac 1{(2\pi i)^2}\int_{\sigma _1-i\infty }^{\sigma
_1+i\infty }e^{p_1\tau }dp_1\int_{\sigma _2-i\infty }^{\sigma _2+i\infty
}E_{p_1p_2}e^{p_2\zeta }dp_2; \\
\frac{\partial E(\zeta ,\tau )}{\partial \zeta } &=&\frac 1{(2\pi i)^2}%
\int_{\sigma _1-i\infty }^{\sigma _1+i\infty }dp_1\int_{\sigma _2-i\infty
}^{\sigma _2+i\infty }dp_2p_2E_{p_1p_2}e^{p_1\tau +p_2\zeta }-\frac{E(0,\tau
)}{2\pi i}\int_{\sigma _2-i\infty }^{\sigma _2+i\infty }\frac{e^{p_2\zeta }}{%
p_2}dp_2,
\end{eqnarray}

\noindent where $\sigma _1,\sigma _2>0$ , and there are introduced
normalized dimensionless variables
\begin{equation}
\zeta =kx;\qquad \tau =t/t_0;\qquad t_0=\frac 1k\sqrt{\frac{m_e}{2k_BT}}=%
\frac 1{kv_T};
\end{equation}

\noindent $k$ is real; $v_T$ is thermal mean velocity. Here appears also
some $E(0,\tau )$ as a boundary field constant of integration. It can be
naturally considered as a given field related to some external circumstances
as it is due to some external sources, etc.

After substitution of eqs. (16), (17) and (18) into eq. (15) one obtains
\begin{eqnarray}
&&-4\pi ^2\alpha _1(\zeta )e^{i\zeta }+A\int_{\sigma _1-i\infty }^{\sigma
_1+i\infty }dp_1\int_{\sigma _2-i\infty }^{\sigma _2+i\infty
}dp_2E_{p_1p_2}e^{p_1\tau +p_2\zeta }\times   \nonumber \\
&&\{\int_0^{\zeta /\tau }d\xi \xi e^{-\xi ^2}[\frac{1-e^{-p_2\xi \tau
-p_1\tau }}{p_1+p_2\xi }-\frac{1-e^{p_2\xi \tau -p_1\tau }}{p_1-p_2\xi }%
]+\int_{\zeta /\tau }^\infty d\xi \xi e^{-\xi ^2}[\frac{1-e^{-p_1\zeta /\xi
-p_2\zeta }}{p_1+p_2\xi }-\frac{1-e^{-p_1\zeta /\xi +p_2\zeta }}{p_1-p_2\xi }%
]\}  \nonumber \\
&=&\int_{\sigma _1-i\infty }^{\sigma _1+i\infty }dp_1\int_{\sigma _2-i\infty
}^{\sigma _2+i\infty }dp_2p_2E_{p_1p_2}e^{p_1\tau +p_2\zeta }-2\pi iE(0,\tau
)\int_{\sigma _2-i\infty }^{\sigma _2+i\infty }\frac{e^{p_2\zeta }}{p_2}dp_2,
\\
&&  \nonumber
\end{eqnarray}

\noindent where $A=const$; stepwise function $\alpha _1$ is proportional to $%
\alpha (x)$.

Keeping in mind the subsequent consideration of some longitudinal analogy of
the skin effect, we shall suppose the field $E(0,\tau )$ takes the form

\begin{equation}
E(0,\tau )=E_0e^{-i\beta \tau },
\end{equation}

\noindent where $\beta $ is some real number.

Taking in account equalities of the type
\begin{equation}
e^{-i\beta \tau }=\frac 1{2\pi i}\int_{\sigma _1-i\infty }^{\sigma
_1+i\infty }\frac{e^{ip_1\tau }}{p_1+i\beta }dp_1,
\end{equation}

\noindent and finiteness of the integrands in points $\xi $ with $p_1\pm
p_2\xi \rightarrow 0$ for terms of the type
\[
\frac{1-e^{\pm (p_1\pm p_2\xi )}}{p_1\pm p_2\xi },
\]

\noindent so one assuming the following integrations along $\xi $ in the
sense of the principal value, one can find an asymptotical solution of eq.
(19) as
\begin{equation}
E_{p_1p_2}=\frac{\frac{\alpha ^{*}}{p_1p_2(p_2-i)}+\frac{E_0}{%
p_2^2(p_1+i\beta )}}{1+\frac{2A\bar{\xi}^2e^{-\bar{\xi}^2}\Delta \xi }{%
p_1^2-p_2^2\bar{\xi}^2}}
\end{equation}

\noindent where $\alpha ^{*}\sim \alpha _0$ , and $\bar{\xi}$ is some
characteristic value of the normalised velocity $\xi \equiv v/v_T;\;\bar{\xi}
$ and $\Delta \xi $ are of the order of $1$, and
\begin{equation}
A\equiv \frac{4\sqrt{\pi }e^2n_e}{k^2k_BT}.
\end{equation}

In the derivation of eq. (22) it is supposed that one may apply to integrals
over $\xi $ in the braces of eq. (19) the mean value theorem, replacing
functions of $\xi $ in the integrands with the functions of some constant
value $\bar{\xi}$ belonging to the interval of integration. Also at
integration over $p_1,p_2$ one can neglect arising terms
\begin{equation}
\frac{e^{-p_2\bar{\xi}\tau -p_1\tau }}{p_1+p_2\bar{\xi}};\qquad \frac{%
e^{-p_1\zeta /\bar{\xi}-p_2\zeta }}{p_1+p_2\bar{\xi}}.
\end{equation}

\noindent asymptotically for $\sigma _1,\sigma _2>0$ and large $\zeta ,\tau
\rightarrow \infty $. At some fixed value of $\sigma _2>0$ Laplace
transformation implies the integration contour $\sigma _1\pm i\infty $ to be
at right of the pole $p_1-p_2\bar{\xi}=0$ $(\sigma _1>\sigma _2\bar{\xi})$;
this also means the possibility to neglect in an asymptotical solution with
large $\zeta ,\tau \rightarrow \infty $ the exponentially small terms
\begin{equation}
\frac{e^{p_2\bar{\xi}\tau -p_1\tau }}{p_1-p_2\bar{\xi}};\qquad \frac{%
e^{-p_1\zeta /\bar{\xi}+p_2\zeta }}{p_1-p_2\bar{\xi}}.
\end{equation}

Asymptotical value of $E(\zeta ,\tau )$ at both $\zeta \rightarrow \infty $
and $\tau \rightarrow \infty $ is then defined by the pair poles of
expression (22) at shifting the integration contours $\sigma _1,\sigma _2$
to the left. To the term with $\alpha _0\neq 0$ there corresponds a
constituent of the asymptotical solution $E(\zeta ,\tau )$, defined by the
residua sum in the poles
\[
(p_1=0,\;p_2=i);\qquad (p_1=0,\;p_2=\pm \sqrt{2Ae^{-\bar{\xi}^2}\Delta \xi }%
);
\]

\begin{equation}
(p_1=\pm i\bar{\xi}\sqrt{2Ae^{-\bar{\xi}^2}\Delta \xi +1},\;p_2=i)
\end{equation}
(the poles $(p_1,p_2)=(0,0)$ do not contribute to $E(\zeta ,\tau )$ since $%
E_{p_1p_2}\rightarrow 0$ at successive transits $p_1\rightarrow 0$, then $%
p_2\rightarrow 0$ or v.v.).

The phase speed of waves, which correspond to the last pair $(p_1,p_2)$ in
(26) at small $n_e$, coincides with the mean thermal velocity
\[
\frac{\omega _0}k\simeq \frac 1{t_0k}=\sqrt{\frac{2k_BT}{m_e}}=v_T,
\]

\noindent what is in accordance with experimental plasma echo speeds. At
large $A$ one finds
\[
\frac{\omega _0}k\simeq \frac{\sqrt{2A}}{t_0k}=\frac{\gamma \omega _L}k%
\simeq \frac{\omega _L}k,
\]

\noindent where $\omega _L$ if Langmuir frequency, and $\gamma $ is a factor
of the order of $1$.

For the term in (22) with $E_0\neq 0$ (with expected extinguishing plane
longitudinal waves) one obtains pole pairs defining additional constituents
of the asymptotical solution:
\[
(p_1=-i\beta ,\;p_2=0);\qquad (p_1=\pm i\bar{\xi}\sqrt{2Ae^{-\bar{\xi}%
^2}\Delta \xi },\;p_2=0);
\]

\begin{equation}
(p_1=-i\beta ,\;p_2=\pm \sqrt{2Ae^{-\bar{\xi}^2}\Delta \xi -(\beta /\bar{\xi}%
)^2}).
\end{equation}

\noindent But in the pole $p_2=0$ of the second order (that is $1/p_2^2$)
the residium equals zero.

The ``mixed'' constituent of solution which corresponds to the poles $%
(p_1=-i\beta ,p_2=i)$ at $x,t\rightarrow \infty $ , gives zero contribution.

The derived asymptotical solutions content some exponentially growing modes.
This is easy explained by the non-selfconsistency of assumed initial and
boundary conditions $f_1(\vec{v},x,0)$ and $E(0,t)$ in the absence of an
outside source field. The divergent terms can be easily removed with the
corresponding fitting amplitude values $\alpha ^{*}$ and $E_0$ in eq. (22)
and assuming $\beta =0$. So one obtains only nondamping standing
oscillations as a sum of modes moving with the Langmuir velocity at large $%
n_e$ and the one-dimensional mean thermal velocity $v_T$ at small $n_e$.

One may suppose that an outside source field might be accounted for by
dividing the boundary value of the field into a part of the selfconsistent
plasma response and the rest one. The first one is defined by condition of
the mutual cancellation of terms, which exponentially grow at $x\rightarrow
\infty $. The second one gives proper plasma modes induced by an outside
field.

The above found very simple result of a non-damping standing wave for above
variant of the self-consistent initial and boundary conditions seems to be
very natural physically.

In a variant of a pulse exitation of longitudinal waves with the field $E$
along $x$ at
\begin{equation}
\alpha (x)=0;\qquad E(0,\tau )=E_0\delta _{+}(\tau )
\end{equation}

\noindent one can obtain with the Laplace transformation of $\delta $%
-function $\delta _{+}(\tau )$ being $1$ (see [6]) a solution
\begin{equation}
E_{p_1p_2}=\frac{E_0/p_2^2}{1+\frac{2A\bar{\xi}^2e^{-\bar{\xi}^2}\Delta \xi
}{p_1^2-p_2^2\bar{\xi}^2}},
\end{equation}

\noindent respectively only two pair poles
\begin{equation}
(p_1=i\bar{\xi}\sqrt{2Ae^{-\bar{\xi}^2}\Delta \xi },\;p_2=0);\qquad (p_1=-i%
\bar{\xi}\sqrt{2Ae^{-\bar{\xi}^2}\Delta \xi },\;p_2=0);
\end{equation}

\noindent with zero residua and asymptotic value $E(x,t)$.

These results, which are in principle accessible to experimental
verifications, are discrepant qualitatively with the Landau theory, so this
means the necessity of a new ideology of plasma waves. The asymptotic limit
is represented by a superposition of a finite number of coupled both damping
and {\sl non-damping} oscillatory modes with different dispersion laws.

Using the method of Laplace transformation one may easily obtain also
asymptotic solution for the perturbation $f_1(\vec{v},x,0)$. In this case an
arising pole $(p_1+p_2\bar{\xi})$ and others may indeed lead among others to
terms in $f_1$ of the type $\exp [ik(x-vt)]$, as in the Landau theory, but,
in contrary to it, these terms are related to {\sl non-damping} electrical
fields.

6. The spreading of transverse electromagnetic waves in a collisionless
plasma is traditionally controlled by the Landau rule of passing around the
pole and then solving an artificially constructed dispersion equation. But
as above, in this case the precise (i.e. without Landau additives)
dispersion equation for the linearized and Maxwell field equations,
resulting after substitution therein the travelling wave exp$(-iwt+ikx)$,
has no solutions.

The substitution into the Maxwell field equation of the solution of the
kinetic equation, which had been obtained with the method of
characteristics, leads in non-relativistic approach (after omitting poles
terms of the order $(v/c)^2$) to the common optical wave equation for a wave
in refractive medium with the refracting index $n=\sqrt{1-\omega _L^2/\omega
_0^2}$, where $\omega _L$ is the electron Langmuir frequency, so the
solution at real $n$ is represented by the travelling wave
\begin{equation}
E_{\perp }\sim \exp (-i\omega _0t+ikx)
\end{equation}

These results are well known, but it is worth to note however that in the
case of medium separating (boundary) surfaces there is possible a peculiar
non-orthodox solution of the optical equation, as a differential equation
with ruptured boundary conditions, different from the waves (31), but that
one is unstable and questionable to be realized experimentally, e.g. in
femtosecond spectroscopy [7]. The pole problems do not arise in this
consideration at all, whatever singularities in the non-relativistic
approach are absent (this corresponds to the well known fact, that there are
large $v$'s in poles, so that the Maxwell distribution function $f_0(v)$
there is very small). The pole problem certainly arises in the relativistic
case and could be solved by applying the above method of two-dimensional
transformation to the relativistic kinetic equation.

As for the experimental verification of the Landau theory, first of all we
note that there is a very small number of such works, which also were
carried out by the same scientists [8,9]. Such experiments in fact should be
very delicate. Certainly, it is very difficult to obtain experimentally a
collisionless plasma with Maxwellian electron distribution function due to
conflicting demands: the Maxwell distribution is just commonly a consequence
of electron energy exchange in collisions, that is the plasma must be
collisional. There is also a possibility in principle of a dependence of
perturbation spreading in plasma on geometric, spectral and other features
of an excitation source.

\vspace{0.5in}

\begin{center}
{\Large Conclusions}
\end{center}

{}From an unprejudiced consideration of the Landau theory of spreading and
damping of plasma waves there is emphasized fundamental fact that the
dispersion equation, derived in accord with the classical canons, does not
have whatever solutions, but which {\sl must} exist at least for the most
slowly damping Landau's asymptotic travelling wave, be it a case of
collisionless or low-collision plasmas. The bypass way, which had been
chosen by Landau in 1945 and is now cited in all textbooks on plasma physics
almost without variations (cf. [4,5]), respectively Landau's rule of passing
around poles, appears to be unsatisfactory because  at tending the imaginary
part $\delta $ of the frequency $\omega $ to zero resulting expressions lose
their sense since they do not tend to any definite limits. Thus Landau's
analytical continuation is really not existent. This is related to the
impossibility (e.g. after integration in the dispersion equation by velocity
spherical angles $d\varphi d\theta $) of expansion in small $\delta /\omega
_0$ (that  results in asymptotically divergent series, cf. [2]). Landau's
theory results in a quite sophisticated and mysterious theory of the plasma
echo, which leads to the paradox of conflicting simultaneous existence of a
non-damping mode of plasma echo relative to the Landau field-damping wave.
At the same time it is possible to construct some reasonable,
non-contradictory and mathematically unequivocal theory of spreading of
plasma perturbations, using a more strict procedure of the two-dimensional
Laplace transformation, which leads asymptotically to a system of coupled
damping and non-damping modes of plasma waves and oscillations, but not to
an asymptotic single travelling wave with a definite dispersion law, as it
should follow from the Landau paradigm.

The rejection of Landau's theory will open many new perspectives in numerous
problems of the plasma physics.

One can explain existing now widely adherence to the Landau theory not only
by some natural conservatism and piety to Landau, but more by the fact of
rather imperceptible, practically unobservable effects arising from this
theory in all surrounding us realities (laboratory, technical, experimental,
observable cosmic, etc.) in spite of the ultimately wide its diffusion in
abstract sophistications of theorists.

\end{document}